\documentclass{PoS}
\usepackage{wrapfig}
\usepackage{multirow}

\newcommand{\eeZHH}{$e^+e^-\rightarrow {ZHH}$}
\newcommand{\eevvHH}{$e^+e^-\rightarrow\nu\bar\nu {HH}$}
\newcommand{\eeZH}{$e^+e^-\rightarrow {ZH}$}
\newcommand{\eevvH}{$e^+e^-\rightarrow\nu\bar\nu {H}$}
\newcommand{\eeeeH}{$e^+e^-\rightarrow e^+e^-{H}$}
\newcommand{\eettH}{$e^+e^-\rightarrow t\bar{t} {H}$}
\newcommand{\Htobb}{$H\to b\bar{b}$}
\newcommand{\HtoWW}{$H\to WW^*$}
\newcommand{\HtoZZ}{$H\to ZZ^*$}
\newcommand{\eeZHtobb}{$e^+e^-\rightarrow {ZH}\to Zb\bar{b}$}
\newcommand{\eevvHtobb}{$e^+e^-\rightarrow\nu\bar\nu {H}\to\nu\bar{\nu}b\bar{b}$}

\title{Measurement of Higgs couplings and self-coupling at the ILC}

\ShortTitle{Measurement of Higgs couplings and self-coupling at the ILC}

\author{\speaker{Junping TIAN}%
        \thanks{On behalf of the ILD concept group.}\\
       KEK\\
       E-mail: \email{junping.tian@kek.jp}}

\author{Keisuke Fujii$^\dagger$\\
        KEK\\
        E-mail: \email{keisuke.fujii@kek.jp}}

\abstract{In the Standard Model (SM) the couplings of the Higgs boson to SM particles and itself (self-couplings) are uniquely specified once the masses of the particles in question as well as the Higgs boson mass are given. Precision measurements of these couplings in the future collider experiments are the key to either verifying the mechanism of the electroweak symmetry breaking in the SM or uncovering physics beyond the SM. This article gives an overview of how various Higgs couplings will be measured at the ILC. Emphasis is put on the ILC's capability of performing fully model independent determination of absolute $HZZ$ and $HWW$ couplings, the Higgs total width, and hence various other Higgs couplings, which cover essentially all the crucial ones including the top-Yukawa coupling $Htt$ and the trilinear Higgs self-coupling $\lambda_{HHH}$. The strategy to get the best precision measurements at the ILC is through staged running, which provides many independent $\sigma\times\mathrm{Br}$ measurements for different production channels and at different energies. A method of global fitting is discussed to utilize all of the available information and to derive combined precisions.}

\FullConference{The European Physical Society Conference on High Energy Physics \\
		 18-24 July, 2013\\
		 Stockholm, Sweden}

\begin{document}

\section{Introduction}
\begin{wrapfigure}{r}{0.4\textwidth}
  \vspace{-15pt}  
  \begin{center}
    \includegraphics[width=0.4\columnwidth]{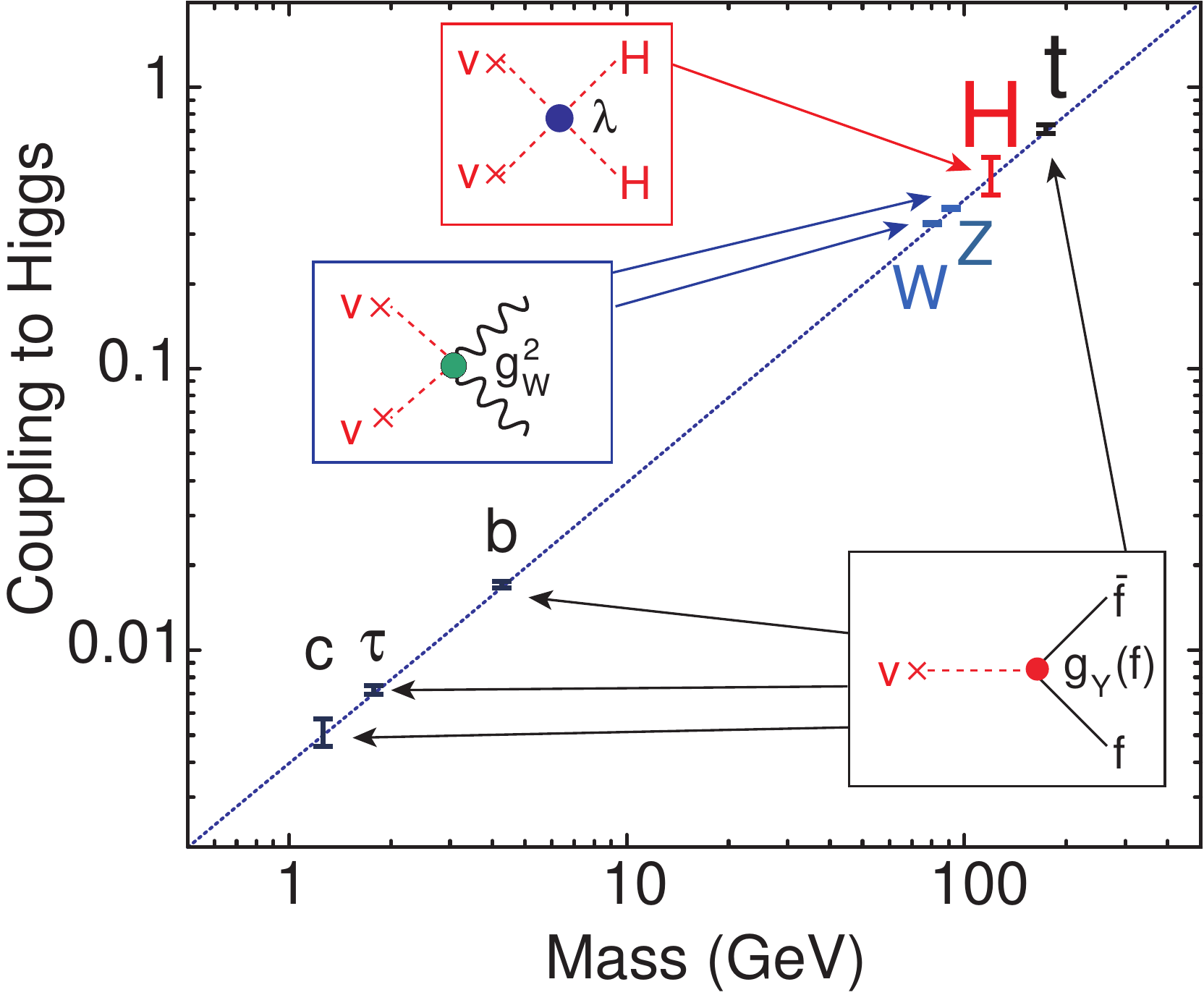}
  \end{center}
  \vspace{-10pt}    
  \caption{Mass-coupling relation \cite{MassCoupling}.}
  \label{fig:MassCoupling}
  \vspace{-5pt}    
\end{wrapfigure}

Based on relativistic quantum field theory and supported by two pillars, gauge principle and electroweak symmetry breaking (EWSB), the Standard Model (SM) has been an extreme success. The first pillar, gauge principle, has been well established by the experiments performed during the past decades at SLC, LEP, Tevatron, and LHC. On the other hand, the second pillar, the mechanism of EWSB, had not been tested at all until the recent discovery of a 125 GeV Higgs(-like) boson \cite{HiggsATLAS,HiggsCMS}. The discovery provides a strong support for the Higgs Mechanism as used in the SM. The mass generation mechanism used in the SM is a minimal solution to the EWSB. However, there is no reason why nature should have chosen this minimal solution. In fact there exist many models of extended Higgs sectors, such as various two Higgs doublet models (2HDM) including its supersymmetric version (SUSY), which provides solutions not only to the EWSB but also to other questions in high energy physics such as the naturalness problem, the dark matter abundance, and the matter and anti-matter asymmetry. The key to verify the SM mass generation mechanism or to open the door to physics beyond the SM (BSM) is to measure the mass-coupling relation as illustrated in Fig.\ref{fig:MassCoupling}, because of the fact that in the SM all the Higgs couplings to SM particles are proportional to their masses. The newly discovered Higgs particle gives us a unique opportunity to test this proportionality, hence to establish the second pillar or to uncover BSM physics. 

The International Linear Collider (ILC), with $\sqrt{s}=200-500~ \mathrm{GeV}$ upgradable to 1 TeV, is exactly the machine to study every detail of the Higgs particle. The technical design report (TDR) \cite{ILCTDRAcc1,ILCTDRAcc2} of the ILC has been just completed. The detailed baseline design (DBD) \cite{ILCTDRDet} of the two proposed detectors, ILD and SiD with a push-pull scheme, has also been completed. The physics reach based on the performance of the designed accelerator and detectors is fully discussed in the physics volume of the TDR \cite{ILCTDRPhys}. This article gives a brief introduction to the measurement of Higgs couplings and self-couplings. For more complete review of the Higgs physics at the ILC, see the ILC Higgs White Paper \cite{WhitePaper} recently prepared for the Snowmass process 2013. 

\begin{figure}[ht]
  \centering
  \begin{tabular}[c]{ccc}
    \includegraphics[width=0.25\columnwidth]{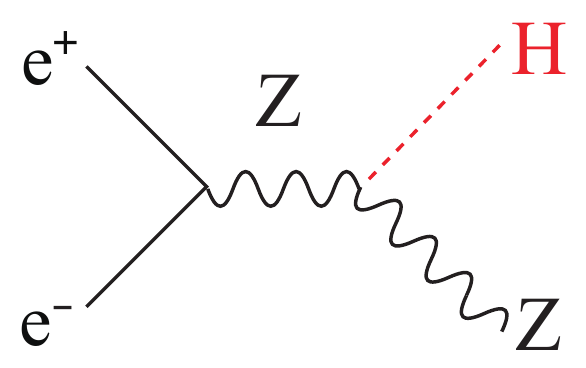}  &
    \includegraphics[width=0.25\columnwidth]{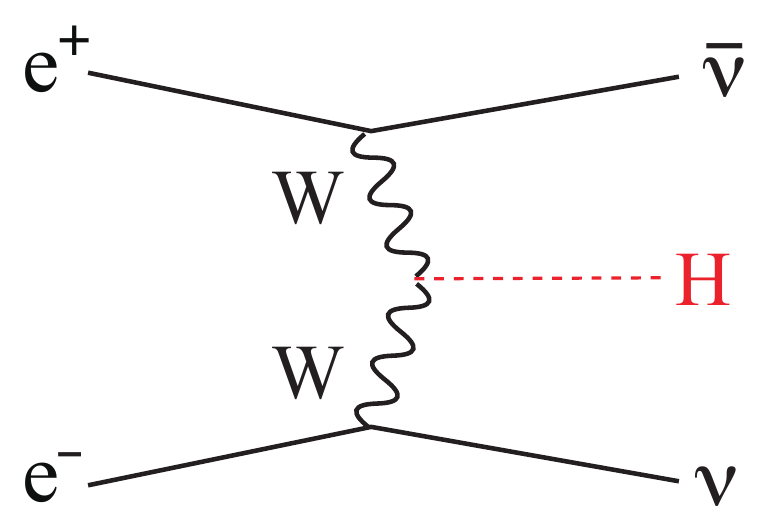}  &
    \includegraphics[width=0.25\columnwidth]{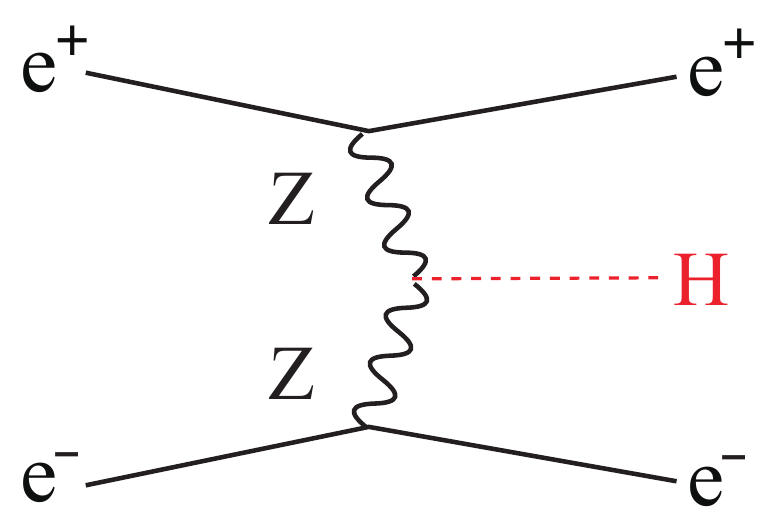}  \\    
 \end{tabular}
  \caption{Major Higgs production processes at the ILC: Higgs-strahlung \eeZH ~(left), $WW$-fusion \eevvH ~(middle) and $ZZ$-fusion \eeeeH ~(right).}
  \label{fig:HiggsProd}
\end{figure}

\begin{figure}[ht]
  \centering
  \begin{tabular}[c]{cc}
    \includegraphics[height=5cm]{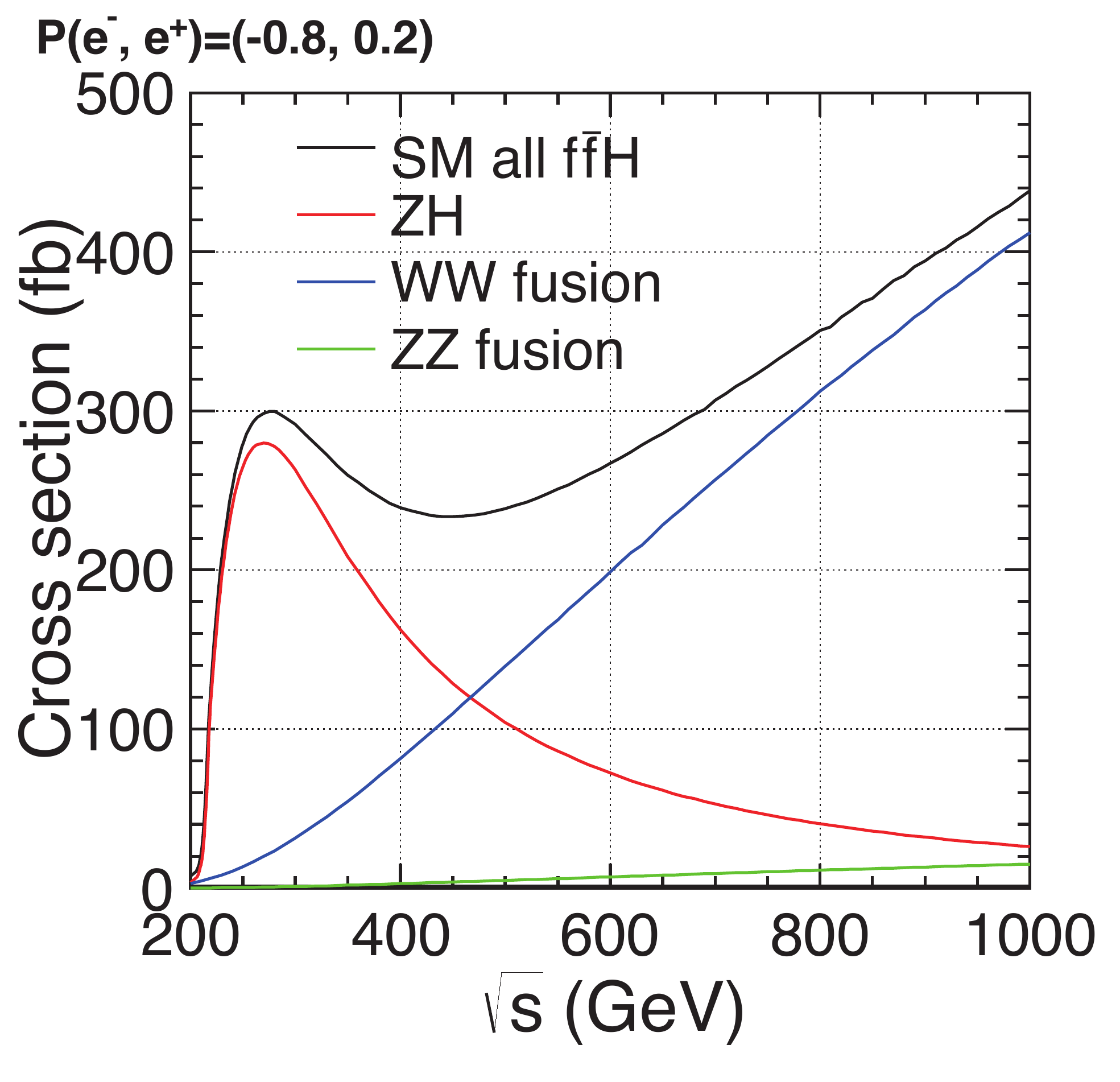}  &
    \includegraphics[height=5cm]{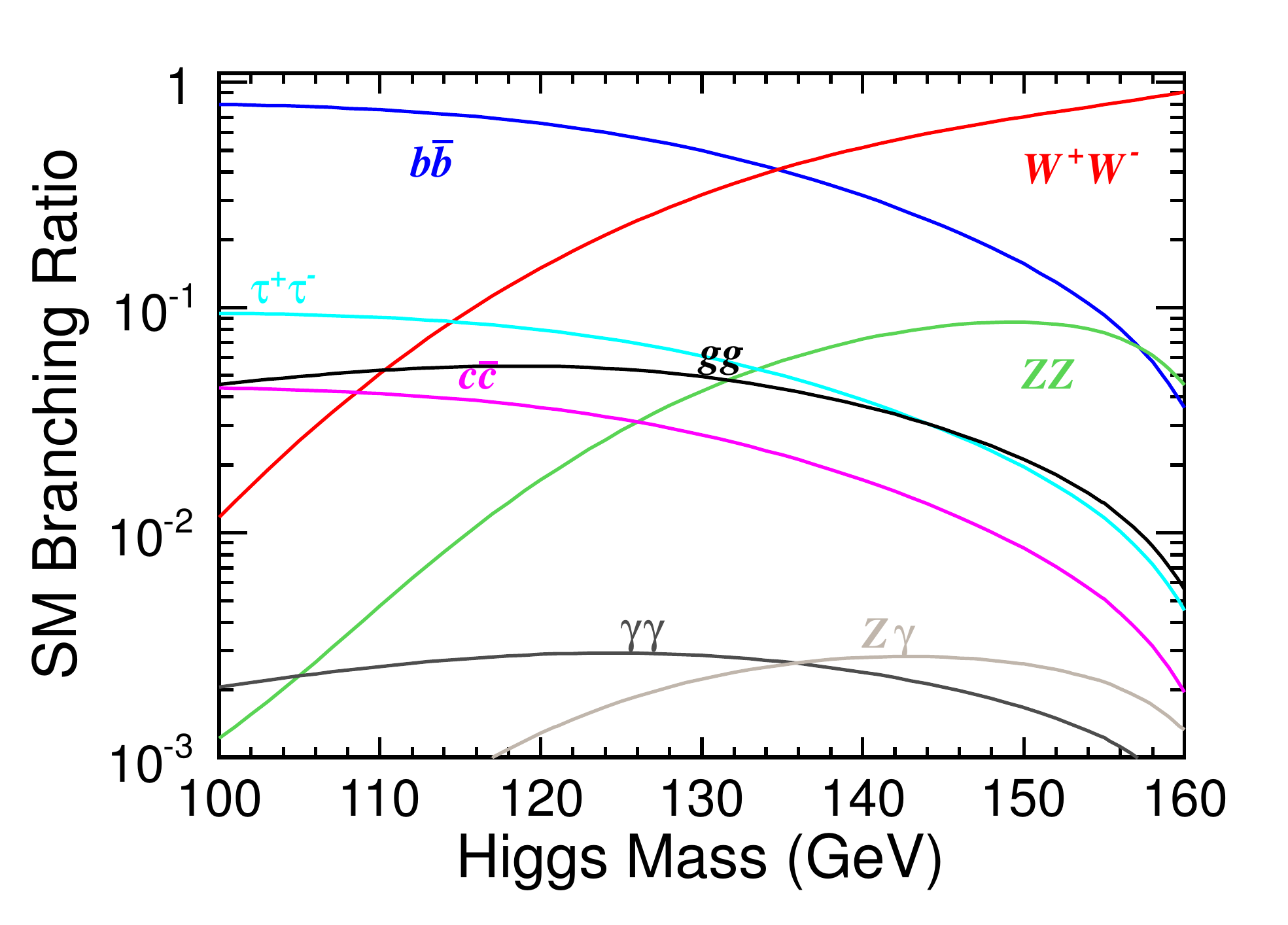}  \\
 \end{tabular}
  \caption{Cross sections of the major Higgs production processes as a function of $\sqrt{s}$ (left) and branching ratios of Higgs decay modes as a function of $m_H$ (right) \cite{ILCTDRPhys}.}
  \label{fig:XSecBRs}
\end{figure}

The major Higgs production processes at the ILC are shown in Fig.\ref{fig:HiggsProd} and their cross sections as a function of $\sqrt{s}$ are shown in Fig.\ref{fig:XSecBRs} (left). Since the Higgs boson decays to $ZZ$ and $WW$ have already been observed at the LHC with rates close to their SM predictions, it is guaranteed that sufficient number of Higgs bosons will be produced at the ILC. The Higgs decay modes and branching ratios as a function of mass are shown in Fig.\ref{fig:XSecBRs} (right).  Notice that , at the Higgs mass of 125 GeV, most of the Higgs decay modes are accessible, which is crucial to get a complete coverage of the mass-coupling relation. A staged running scenario is proposed and the integrated luminosities assumed here are 250 $\mathrm{fb}^{-1}$ at 250 GeV, 500 $\mathrm{fb}^{-1}$ at 500 GeV, and 1000 $\mathrm{fb}^{-1}$ at 1 TeV, which are the nominal values in the baseline design. There is also feasibility of luminosity upgrade \cite{WhitePaper}, which provides 1150 $\mathrm{fb}^{-1}$ at 250 GeV, 1600 $\mathrm{fb}^{-1}$ at 500 GeV, and 2500 $\mathrm{fb}^{-1}$ at 1 TeV. According to the TDR, the electron beam can be polarized to 80\% at all energies; the positron beam can be polarized to 30\% at $250-500$ GeV and to $20\%$ at 1 TeV. The beam polarization is a powerful tool to enhance the cross section of Higgs production. Here $P(e^-,e^+)=(-80\%,+30\%)$ is assumed at $250-500$ GeV, and $P(e^-,e^+)=(-80\%,+20\%)$ is assumed at 1 TeV. The motivation of running energies has been discussed in reference \cite{ILCRunning}.

\section{$HZZ$ coupling}

\begin{wrapfigure}{l}{0.4\textwidth}
  \vspace{-35pt}
  \begin{center}
    \includegraphics[width=0.4\columnwidth]{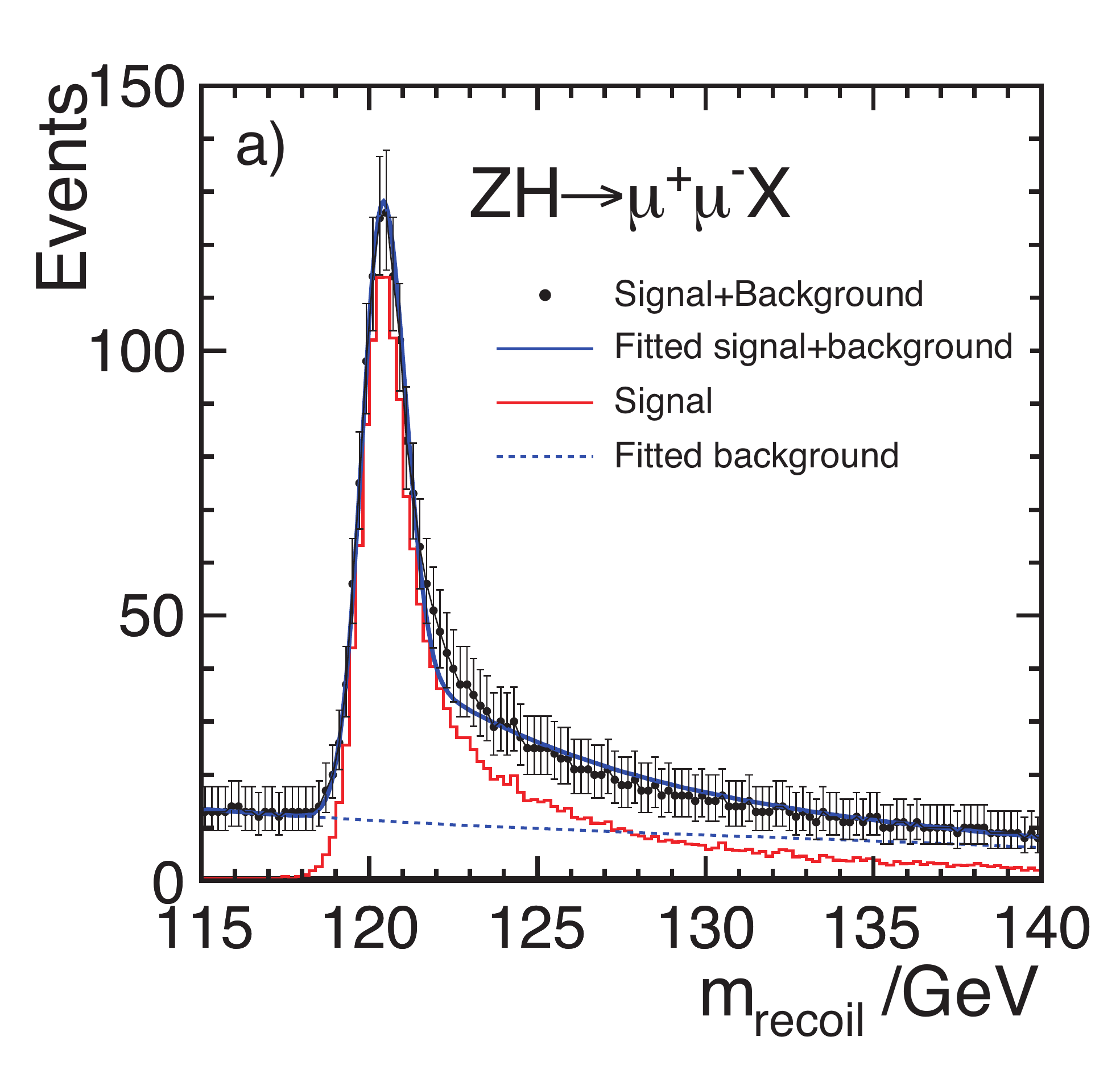}
  \end{center}
  \vspace{-20pt}  
  \caption{Recoil mass distribution of \eeZH ~followed by $Z\to\mu^+\mu^-$ at 250 GeV with a Higgs mass of 120 GeV.}
  \label{fig:RecoilMass}
  \vspace{-20pt}    
\end{wrapfigure}

The well defined four-momentum of initial state ($p_{CM}$) at the ILC allows an inclusive measurement of the cross section of \eeZH. No matter how Higgs decays, as long as we can reconstruct the four-momentum of $Z$ boson ($p_Z$), the Higgs invariant mass ($M_X$) can be calculated by the recoil technique, $M_X=\sqrt{p^2_{CM}-p^2_Z}$. The most effective and precise way to reconstruct $p_Z$ is using the leptonic decay of $Z\to l^+l^-$, $l=e~\mathrm{or}~\mu$. Experimentally we only need to find a lepton pair with an invariant mass consistent with the $Z$ mass and then the recoil mass of the lepton pair is given by $M_X=\sqrt{p^2_{CM}-(p_{l^+}+p_{l^-})^2}$, as shown in Fig.\ref{fig:RecoilMass} \cite{RecoilMass}. By fitting the recoil mass distribution with signal and background components, we get the total cross section of \eeZH ~($\sigma_{ZH}$) since the branching ratios of $Z\to l^+l^-$ are precisely known. From the measured $\sigma_{ZH}$, the coupling $g_{HZZ}$ can be extracted fully model-independently based on $Y_1:=\sigma_{ZH}=F_1g^2_{HZZ}$, where factor $F_1$ can be calculated unambiguously for the Feynman diagram in Fig.\ref{fig:HiggsProd} (left), which is $$g_{HZZ}=\sqrt{\frac{Y_1}{F_1}}.$$ The expected precisions of $g_{HZZ}$ for $m_H=125$ GeV at $\sqrt{s}=$ 250 GeV and 500 GeV are shown in Table \ref{tab:HiggsCouplingsAll}. Obviously the limiting factor here is the momentum resolution for the leptons, which is better at 250 GeV than that at higher energies. 

An alternative approach to reconstruct $p_Z$ is using hadronic decay of $Z\to q\bar{q}$. However, at 250 GeV, $Z$ and $H$ are produced almost at rest. It is not easy to clearly separate the two jets from the $Z$ decay from those from  the Higgs decay. This situation becomes better at 500 GeV where both the $Z$ and $H$ bosons are sufficiently boosted, making their decay products better separated. Thanks to the much larger branching ratio of $Z\to q\bar{q}$, this approach is comparable with the $Z\to l^+l^-$ mode at 500 GeV \cite{WhitePaper}. Nevertheless, more careful study is needed for this approach to investigate the model independence which is not quite clear at this moment.

Overall, the recoil mass measurement using $Z\to l^+l^-$ is one of the most important measurements which gives the absolute $HZZ$ coupling fully model independently. And clearly the best energy for this measurement is 250 GeV, where $\sigma_{ZH}$ reaches its maximum and the momentum resolution is the highest. It's worth mentioning that this recoil mass analysis at 250 GeV also gives the precision Higgs mass measurement, with $\Delta m_H\sim 30 ~\mathrm{MeV}$.

\section{$HWW$ coupling}
\begin{figure}[ht]
  \centering
  \begin{tabular}[c]{cc}
    \includegraphics[height=4.5cm]{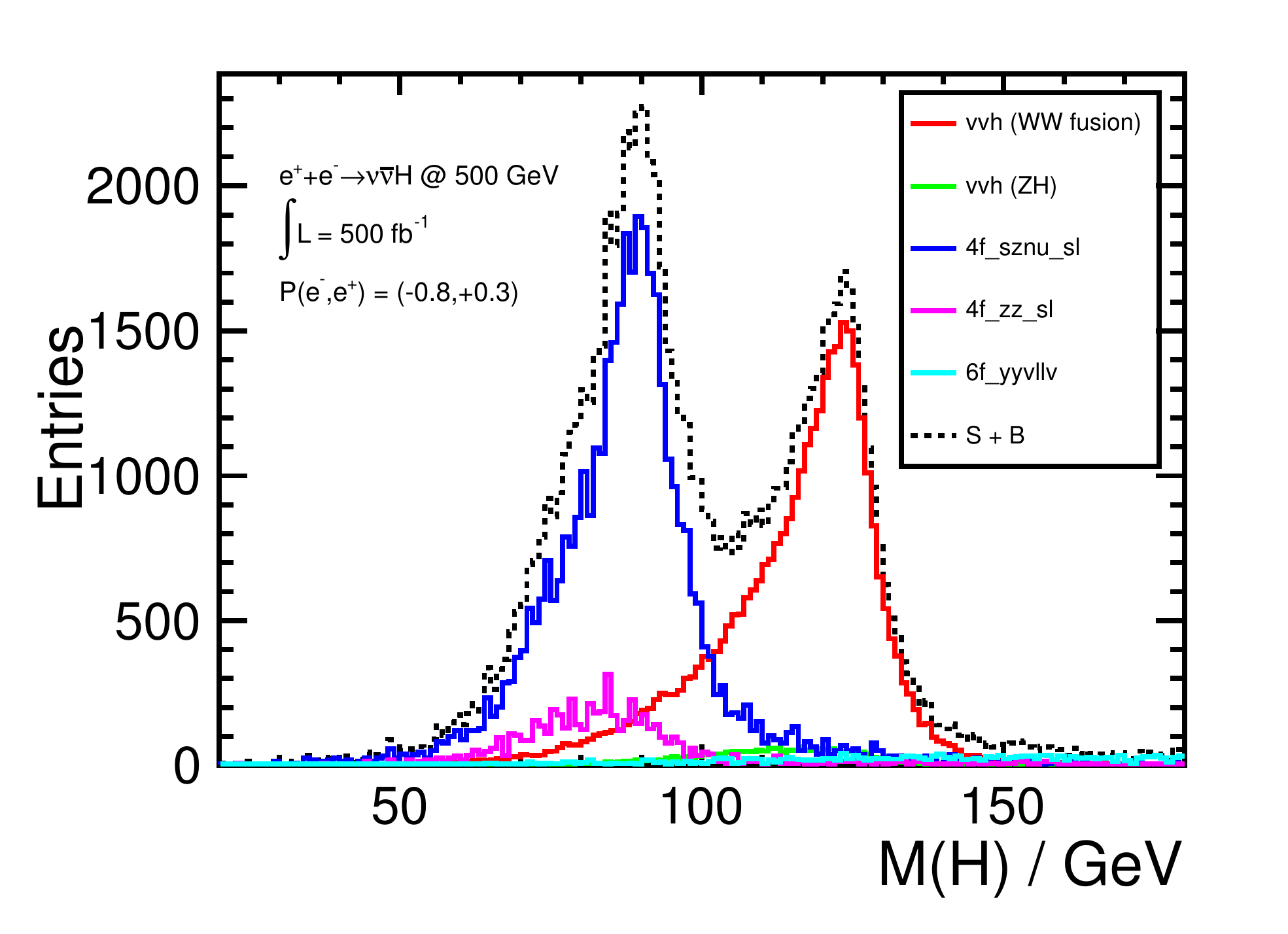}  &
    \includegraphics[height=4.5cm]{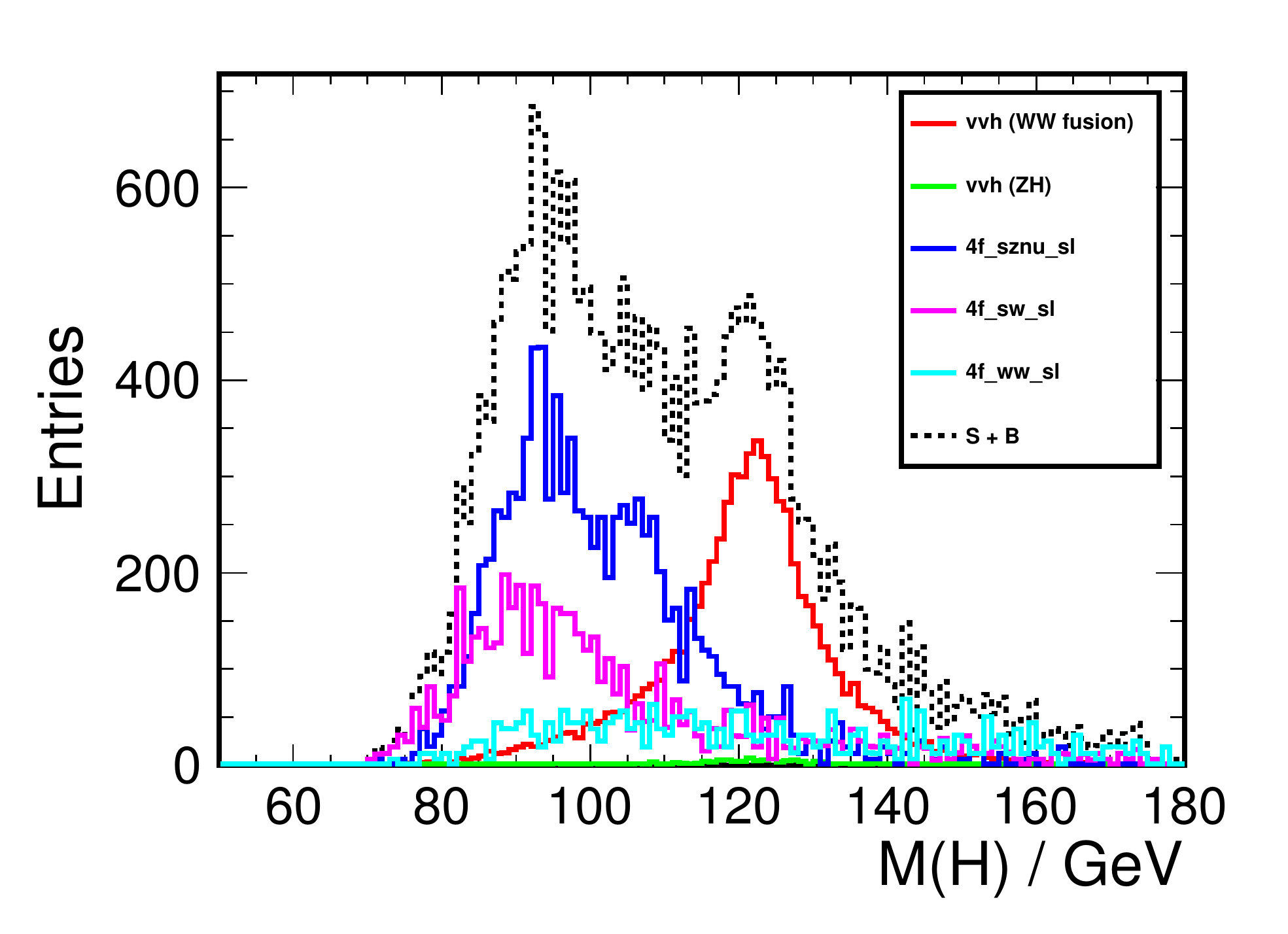}  \\
 \end{tabular}
  \caption{Reconstructed invariant mass of Higgs in the analysis of \eevvH, followed by \Htobb ~(left) and \HtoWW ~(right).}
  \label{fig:vvHbbWW500}
\end{figure}

The $WW$-fusion process \eevvH ~is employed to measure the $HWW$ coupling ($g_{HWW}$), because its cross section $\sigma_{\nu\bar{\nu}H}$ is proportional to $g^2_{HWW}$. Unlike the case of \eeZH ~the final state neutrinos cannot be directly reconstructed by the detector and hence the recoil technique is not applicable any more. We hence need to rely on some exclusive Higgs decay. Here we take advantage of the decay mode with the largest branching ratio: \eevvHtobb ~$\sim 57.8\%$, where the relevant observable is $Y_2:=\sigma_{\nu\bar{\nu}H}\times\mathrm{Br}(H\to b\bar{b})=F_2\cdot g^2_{HWW}\cdot\mathrm{Br}(H\to b\bar{b})$. To extract $g_{HWW}$, we first need to know $\mathrm{Br}(H\to b\bar{b})$. This is measured using the process \eeZHtobb ~\cite{ZHtobb}, where the observable is $Y_3:=\sigma_{ZH}\times\mathrm{Br}(H\to b\bar{b})=F_3\cdot g^2_{HZZ}\cdot\mathrm{Br}(H\to b\bar{b})$. From the ratio of $Y_2/Y_3$, we can obtain the ratio of $g_{HWW}/g_{HZZ}$ and further extract $g_{HWW}$, which is $$g_{HWW}=\sqrt{\frac{Y_1Y_2}{Y_3}\frac{F_3}{F_1F_2}}.$$ The observables $Y_1$ and $Y_2$ based on the Higgs-strahlung process are well measured at 250 GeV. However, $Y_2$ based on the $WW$-fusion process is limited by its small cross section $\sim 14 \mathrm{fb}$ at 250 GeV as shown in Fig.\ref{fig:XSecBRs} (left). At 500 GeV, the cross section of the $WW$-fusion process is one order-of-magnitude larger $\sim 149 \mathrm{fb}$, which allows a precision measurement of $Y_2$. Figure \ref{fig:vvHbbWW500} (left) shows the reconstructed invariant mass of the Higgs candidates in the analysis of \eevvH, followed by \Htobb ~at 500 GeV \cite{vvHbbWW}. The expected precisions of $g_{HWW}$ at $\sqrt{s}=$ 250 GeV and 500 GeV are shown in Table \ref{tab:HiggsCouplingsAll}. Conclusion here is that the absolute $HWW$ coupling can be extracted model independently and going to higher energy is crucial to get the precision of $g_{HWW}$ as good as $g_{HZZ}$.

\section{Higgs total width $\Gamma_H$}
One of the greatest advantages at the ILC is that the Higgs total width ($\Gamma_H$) can be determined model independently by the partial width of the Higgs to $ZZ^*$ ($\Gamma_{H\to ZZ^*}$) decay divided by the branching ratio of \HtoZZ, $\Gamma_H=\frac{\Gamma_{H\to ZZ^*}}{\mathrm{Br}(H\to ZZ^*)}$. The partial width $\Gamma_{H\to ZZ^*}$ can be calculated with the well measured $g_{HZZ}$. But $\mathrm{Br}(H\to ZZ^*)$ is rather statistically limited by its small branching ratio $\sim 2.7\%$. Since $\mathrm{Br}(H\to WW^*)$ is much larger, $\sim 22\%$, a better way is correspondingly by $\Gamma_H=\frac{\Gamma_{H\to WW^*}}{\mathrm{Br}(H\to WW^*)}$, where $\Gamma_{H\to WW^*}$ can be calculated with $g_{HWW}$. An interesting observable here is $Y_4:=\sigma_{\nu\bar{\nu}H}\times\mathrm{Br}(H\to WW^*)=F_4\cdot\frac{g^4_{HWW}}{\Gamma_H}$, which together with known $g_{HWW}$ gives, $$\Gamma_H=\frac{F_4}{Y_4}\cdot g^4_{HWW}=\frac{Y_1^2Y_2^2}{Y_3^2Y_4}\frac{F_3^2F_4}{F_1^2F_2^2}.$$ Figure \ref{fig:vvHbbWW500} (right) shows the reconstructed invariant mass of the Higgs candidates in the analysis of \eevvH, followed by \HtoWW ~at 500 GeV \cite{vvHbbWW}. Obviously for better determination of the Higgs total width one needs higher energy. It is also worth emphasizing that eventually the precisions of $2\frac{\Delta Y_1}{Y_1}$ and $\frac{\Delta Y_4}{Y_4}$ will limit the precision of the total width, since $Y_3$  usually is better measured than $Y_1$, and $Y_2$ is twice better than $Y_4$. The expected precisions of $\Gamma_H$ at different energies are shown in Table \ref{tab:HiggsCouplingsAll}.

\section{Top-Yukawa coupling $Htt$}
\begin{figure}[ht]
  \centering
  \begin{tabular}[c]{cc}
    \includegraphics[height=4.5cm]{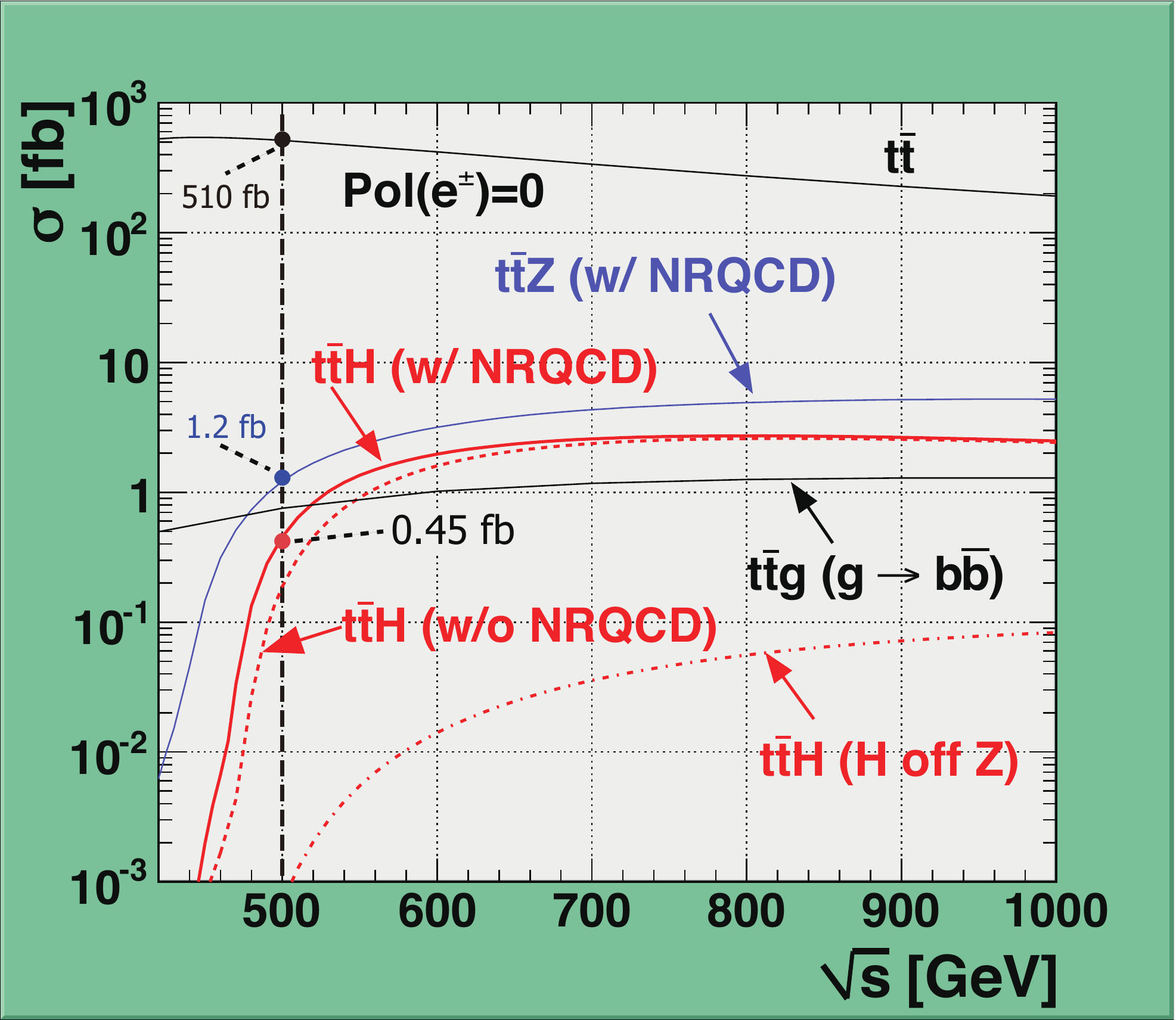}  &
    \includegraphics[height=4.5cm]{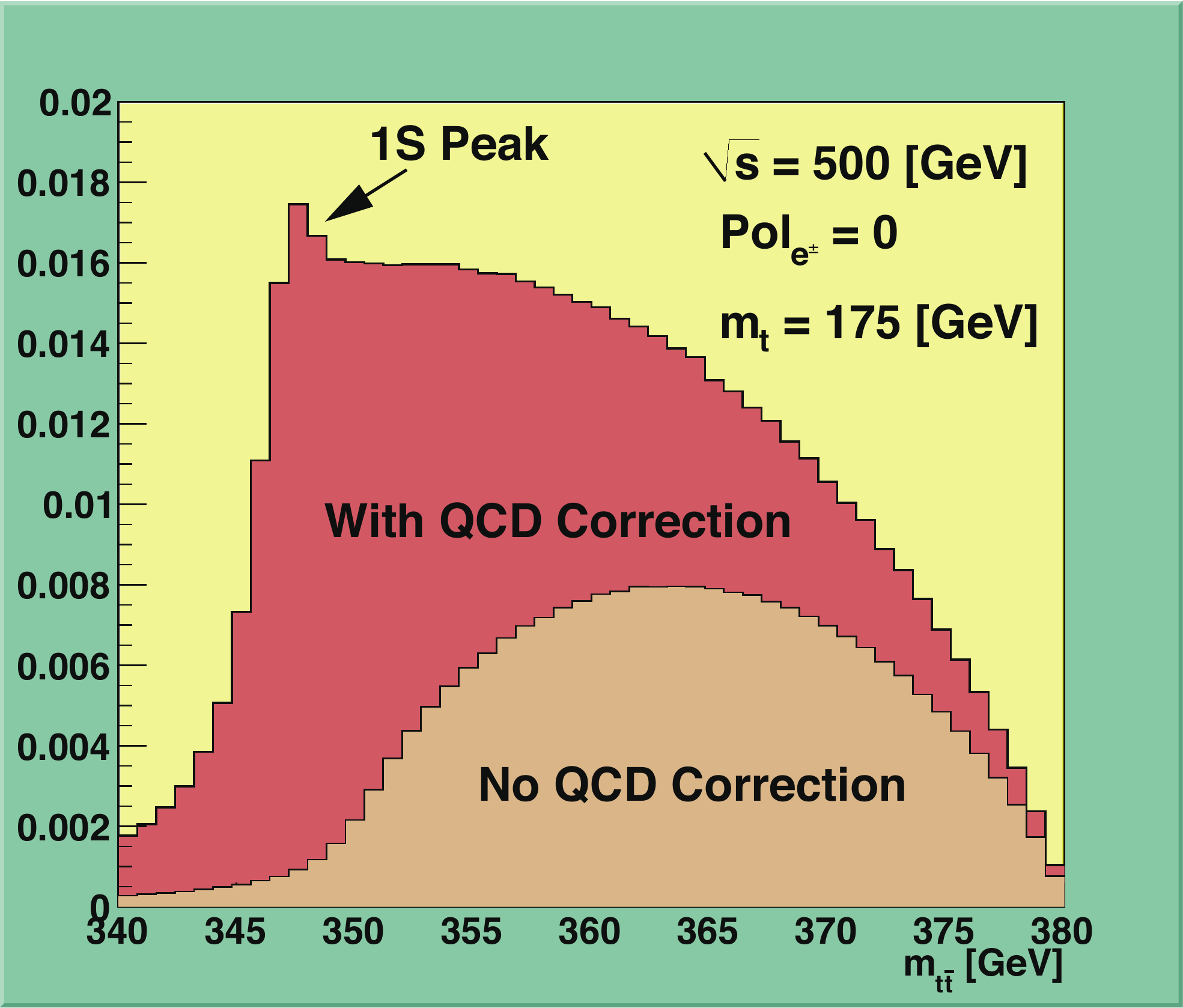}  \\
 \end{tabular}
  \caption{Cross sections for the signal \eettH ~process with and without the non-relativistic QCD (NRQCD) correction together with those for the background processes: $t\bar{t}Z$, $t\bar{t}g(g\to b\bar{b})$ and $t\bar{t}$ (left). The invariant mass distribution for the $t\bar{t}$ subsystem with and without the NRQCD correction (right).}
  \label{fig:ttH}
\end{figure}

The largest Yukawa coupling, top-Yukawa coupling, can be directly accessed at the ILC through the process \eettH. Figure \ref{fig:ttH} (left) shows the cross section of \eettH ~as a function of $\sqrt{s}$ \cite{TopYukawa500}. Though the maximum cross section is reached at round 800 GeV, the cross section at 500 GeV is enhanced by a factor of 2 due to the QCD bound-state effect (non-relativistic QCD correction) as shown in figure \ref{fig:ttH} (right), which makes the measurement of top-Yukawa coupling possible at 500 GeV. The analyses \cite{TopYukawa500,TopYukawa1000} give the expected precisions of the top-Yukawa coupling at 500 GeV and 1 TeV as shown in Table \ref{tab:HiggsCouplingsAll}.

\section{Higgs self-coupling $\lambda_{HHH}$}
\begin{figure}[ht]
  \centering
  \begin{tabular}[c]{cc}
    \includegraphics[height=4.5cm]{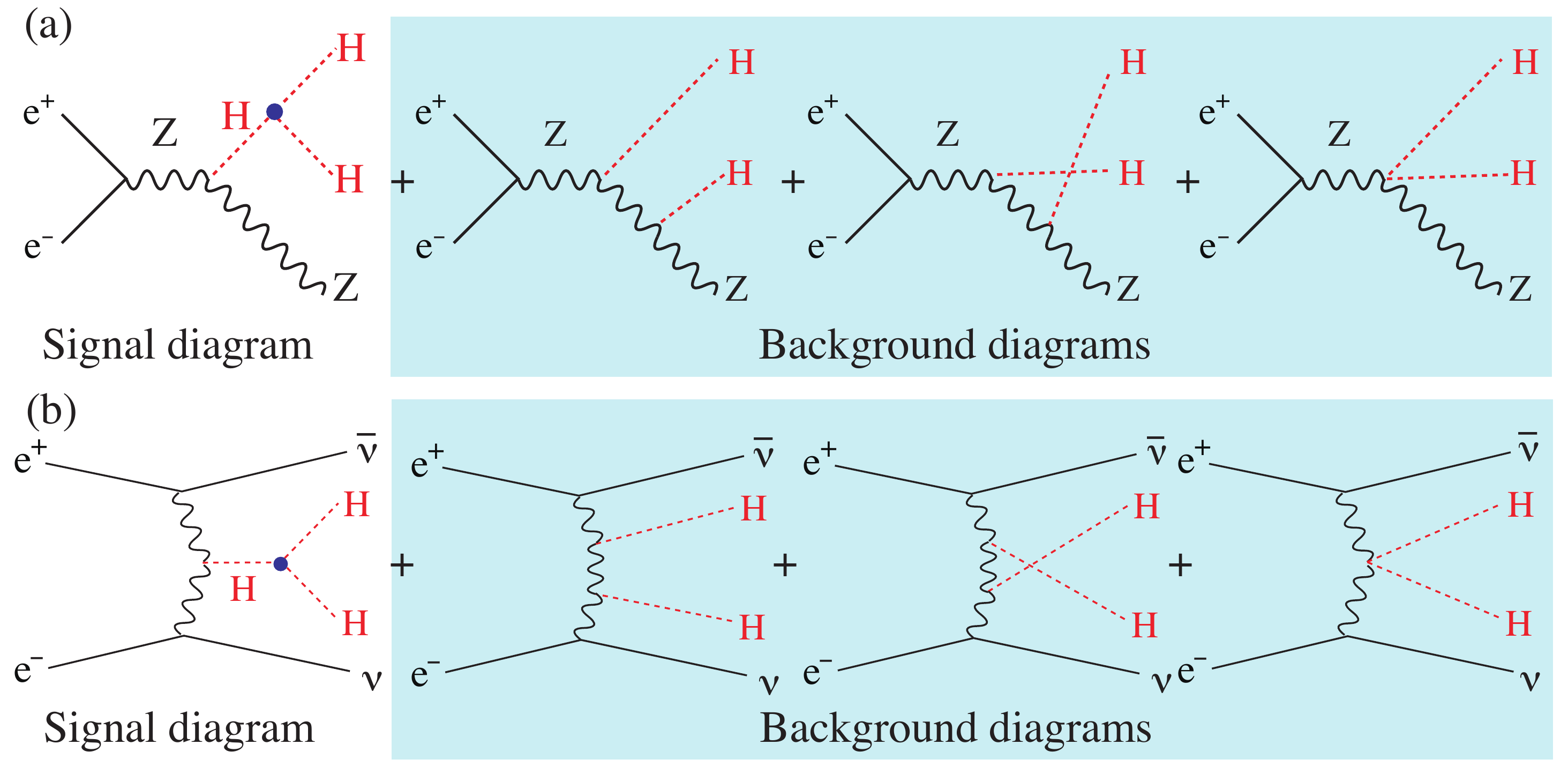} &
    \includegraphics[height=4.5cm]{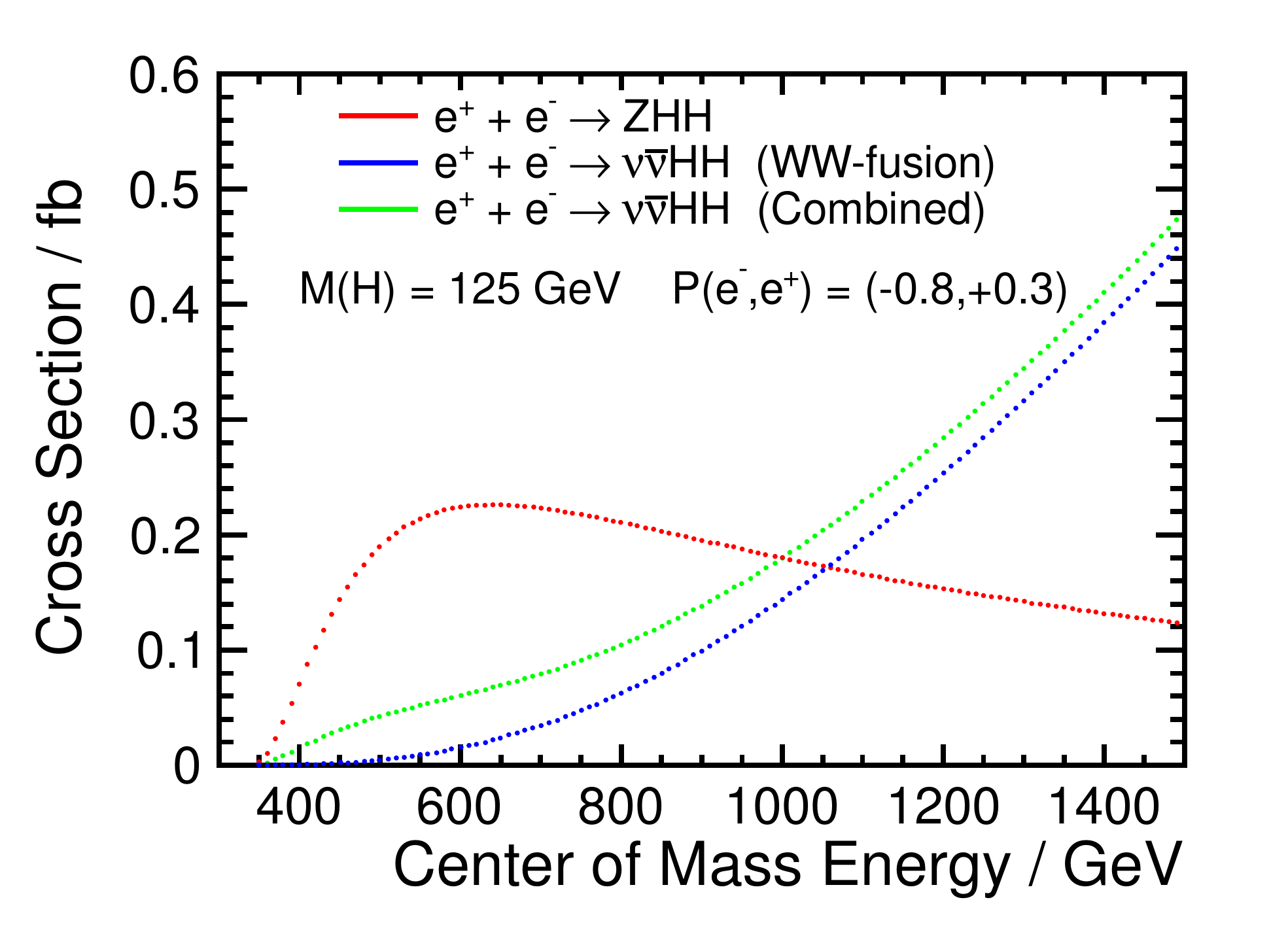} \\
 \end{tabular}
  \caption{Diagrams of the double Higgs production processes, \eeZHH ~(left a) and \eevvHH ~(left b), and their cross sections as function of $\sqrt{s}$ (right).}
  \label{fig:HHProd}
\end{figure}

\begin{figure}[ht]
  \centering
  \begin{tabular}[c]{cc}
    \includegraphics[height=4.5cm]{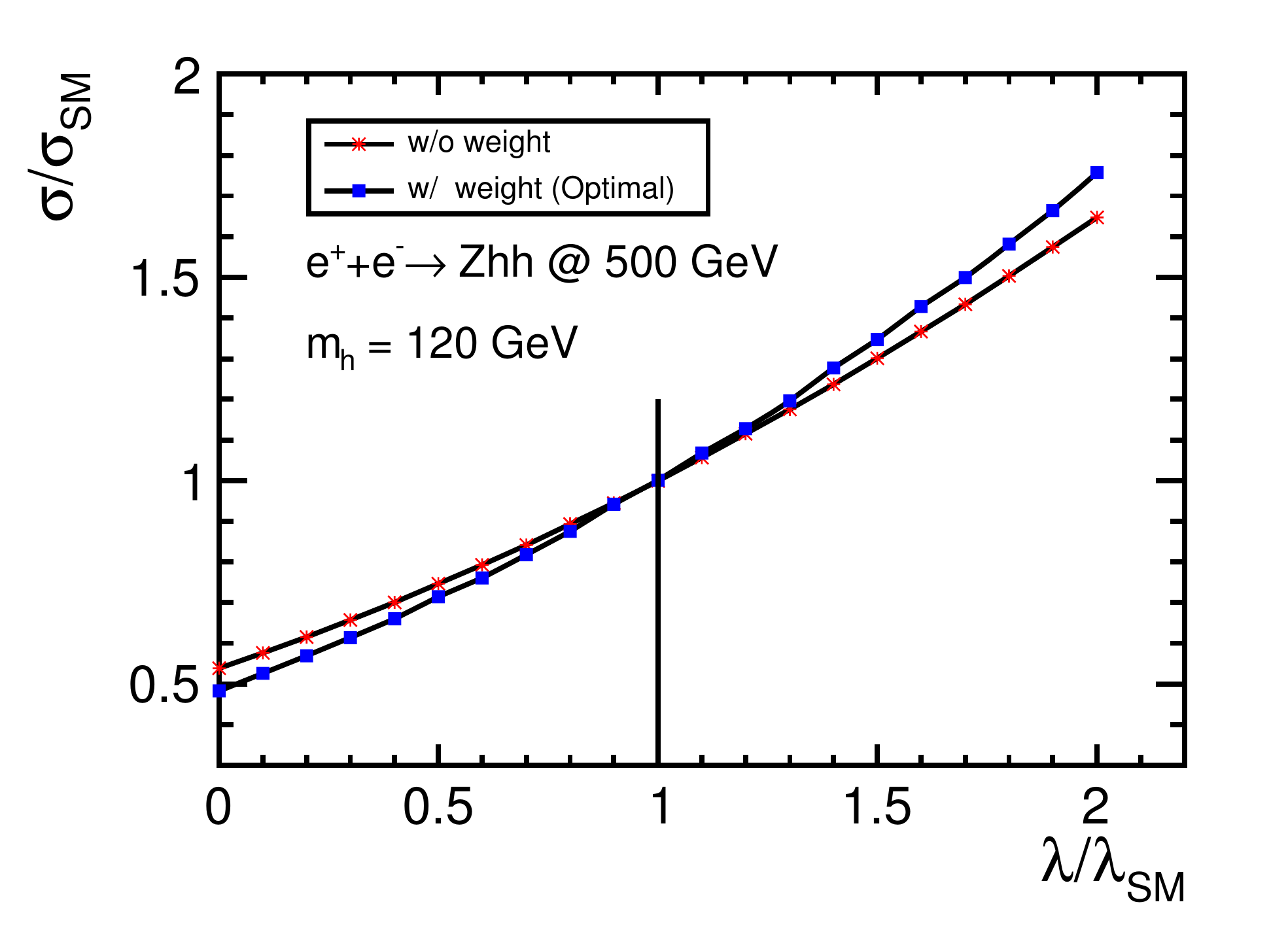} &
    \includegraphics[height=4.5cm]{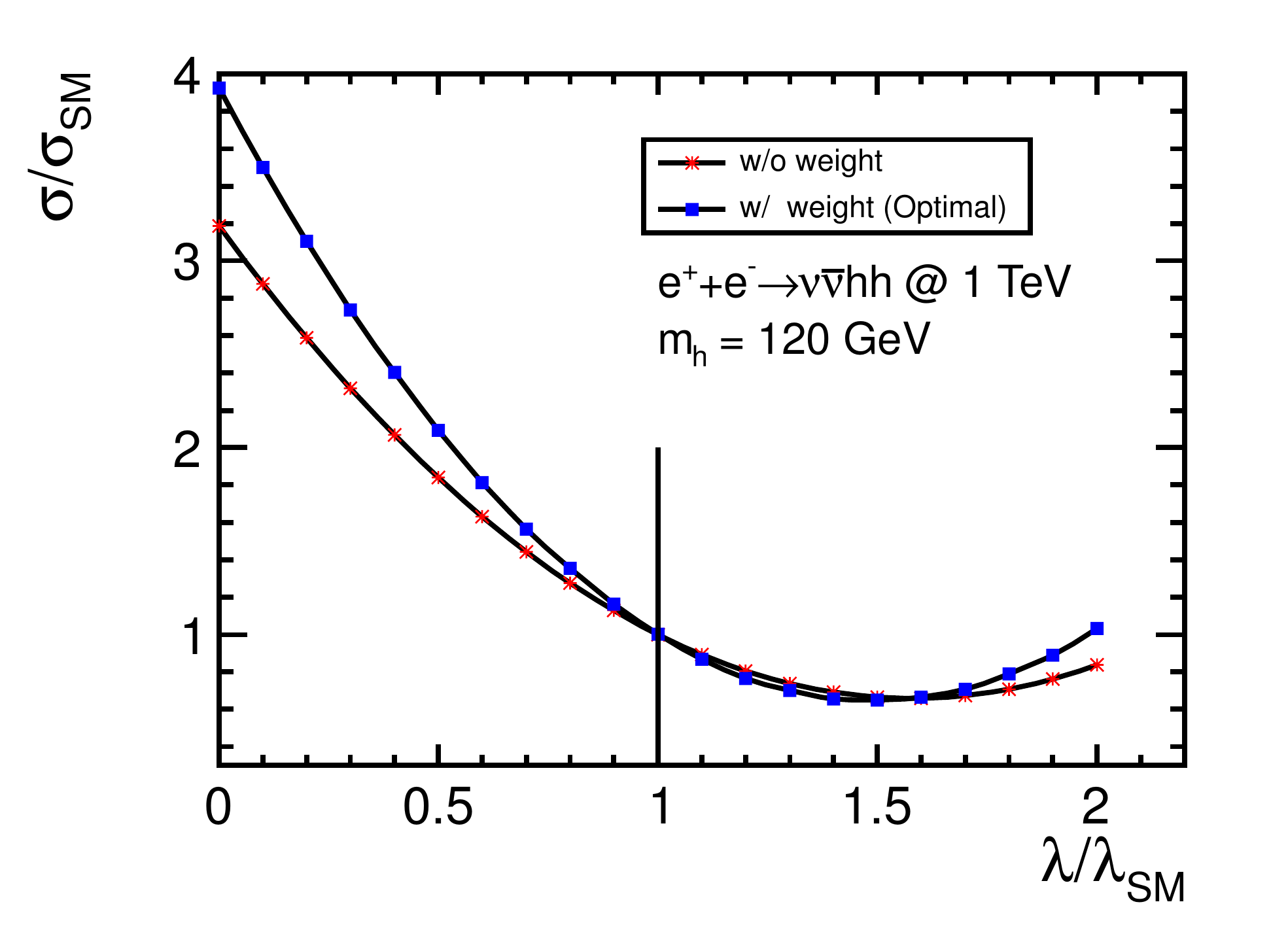} \\
 \end{tabular}
  \caption{Cross sections of \eeZHH ~(left) and \eevvHH ~(right) as a function of Higgs self-coupling.}
  \label{fig:Sensitivity}
\end{figure}

The non-vanishing vacuum expectation value (VEV) of the Higgs field breaks the electroweak symmetry. To fully verify this mechanism, it is necessary to experimentally observe the force that makes the Higgs field condense in the vacuum and measure its strength. The study of the self-coupling can be performed using double Higgs production processes at the ILC, which are shown in Fig.\ref{fig:HHProd} (left). The cross sections of these processes are shown in Fig.\ref{fig:HHProd} (right), where the \eeZHH ~process dominates at around 500 GeV and the WW-fusion process \eevvHH ~becomes very important at higher energies. Nevertheless, the cross sections of both processes are rather small, which makes the study very challenging. In both processes, only the first one is the signal diagram which involves the Higgs trilinear self-coupling $\lambda$; other three ones are irreducible SM background diagrams. This fact leads to that the total cross section $\sigma$ is no longer proportional to $\lambda^2$, which is given instead by $\sigma=a\lambda^2+b\lambda+c$, where the coefficients $a$, $b$, and $c$ are respectively from the signal diagram, the interference, and the background diagrams. Figure \ref{fig:Sensitivity} gives the cross section as a function of self-coupling, with both normalized to SM values. As a result, the relative error on the coupling is no longer half of the relative error on the measured cross section as you would expect if there were only signal diagram, but is $\frac{\delta\lambda}{\lambda}=1.8\frac{\delta\sigma}{\sigma}$ for the $ZHH$ process, and $\frac{\delta\lambda}{\lambda}=0.85\frac{\delta\sigma}{\sigma}$ for the $WW$-fusion process $\nu\bar{\nu}HH$ instead at $\lambda$ equals to its SM value. Using a weighting method \cite{HiggsSelfCoupling1} which weights each event according to its invariant mass of the two Higgs bosons, the cross section becomes more sensitive to the coupling as shown in Fig. \ref{fig:Sensitivity} (red one), the relative error on the coupling can be improved to $\frac{\delta\lambda}{\lambda}=1.66\frac{\delta\sigma}{\sigma}$ for the $ZHH$ process, and $\frac{\delta\lambda}{\lambda}=0.76\frac{\delta\sigma}{\sigma}$ for the $\nu\bar{\nu}HH$ process. Analyses using $HH\to b\bar{b}b\bar{b}$ \cite{HiggsSelfCoupling1} and $HH\to b\bar{b}WW^*$ \cite{HiggsSelfCoupling2} have been carried out based on full detector simulation. Expected precisions of the Higgs self-coupling at 500 GeV and 1 TeV are shown in Table \ref{tab:HiggsCouplingsAll}.

\section{Global fit}
Staged running and various production processes provide many independent $Y_i=\sigma\times\mathrm{Br}(H\to XX)$ measurements with error $\Delta Y_i$ \cite{WhitePaper}, each of which can be predicted by $Y_i^\prime=F_i\cdot\frac{g_{HZZ}^2g_{HXX}^2}{\Gamma_H}$, or $Y_i^\prime=F_i\cdot\frac{g_{HWW}^2g_{HXX}^2}{\Gamma_H}$, or $Y_i^\prime=F_i\cdot\frac{g_{Htt}^2g_{HXX}^2}{\Gamma_H}$, where $XX$ means some specific Higgs decay mode and $F_i$ is some calculable factor corresponding to the search mode. In addition, the recoil mass measurements provide absolute cross section measurements of the \eeZH, process, which can be predicted as $Y_{j}^\prime=F_{j}\cdot g^2_{HZZ}$. To combine all of these measurements to exact the 9 couplings, $HZZ$, $HWW$, $Hbb$, $Hcc$, $Hgg$, $H\tau\tau$, $H\mu\mu$, $Htt$, and $H\gamma\gamma$, and the Higgs total width, $\Gamma_H$, a method of model independent global fit is applied by constructing a $\chi^2$ which is defined as following: $$\chi^2=\sum_{i=1}^{i=N}(\frac{Y_i-Y_i^\prime}{\Delta Y_i})^2,$$ where $Y_i$ is the measured value, $\Delta Y_i$ is the error on $Y_i$, $N$ is the total number of measurements and $Y_i^\prime$ is the predicted value which can always be parameterized by couplings and Higgs total width. Next step is to minimize this $\chi^2$ and get the fitted values of the 10 parameters and their errors. Here we assume all the 9 couplings and the Higgs total width are free parameters without any correlation. The result of our global fit is given in Table \ref{tab:HiggsCouplingsAll}, at different energies and for both baseline and luminosity upgraded scenarios. The systematic errors and theoretical errors are not considered here, which however will be well controlled to below sub-percent level at the ILC.

\begin{table}[t]
 \begin{center}
 \begin{tabular}{|l|r|r|r|r|r|r|}
  \hline
  \multirow{2}{*}{${\Delta g}/{g}$} & \multicolumn{3}{c|}{Baseline} & \multicolumn{3}{c|}{LumiUP} \\ \cline{2-7}
                                                         & 250 GeV & + 500 GeV  & + 1 TeV & 250 GeV & + 500 GeV & + 1 TeV \\
   \hline
   $g_{HZZ}$                                        & 1.3\%       & 1.0\%       & 1.0\%          & 0.61\%   & 0.51\%      &  0.51\% \\
   $g_{HWW}$                                    & 4.8\%       & 1.2\%       & 1.1\%          & 2.3\%     & 0.58\%      &  0.56\% \\   
   $g_{Hbb}$                                       & 5.3\%       & 1.6\%       & 1.3\%          & 2.5\%     & 0.83\%      &  0.66\% \\   
   $g_{Hcc}$                                       & 6.8\%       & 2.8\%       & 1.8\%          & 3.2\%     & 1.5\%        &  1.0\% \\      
   $g_{Hgg}$                                       & 6.4\%       & 2.3\%       & 1.6\%          & 3.0\%     & 1.2\%        &  0.87\% \\      
   $g_{H\tau\tau}$                           & 5.7\%       & 2.3\%       & 1.7\%          & 2.7\%     & 1.2\%        &  0.93\% \\         
   $g_{H\gamma\gamma}$         & 18\%        & 8.4\%        & 4.0\%          & 8.2\%     & 4.5\%        &  2.4\% \\         
   $g_{H\mu\mu}$                            & -                & -                 & 16\%           & -               & -                 &  10\% \\            
   $g_{Htt}$                                          & -                & 14\%        & 3.1\%          & -               & 7.8\%        &  1.9\% \\            
   $\Gamma_H$                               & 11\%        & 5.0\%          & 4.6\%          & 5.4\%      & 2.5\%        &  2.3\% \\            
   $\lambda_{HHH}$                        & -                & 83\%        & 21\%           & -                & 46\%         &  13\% \\                  
   \hline
  \end{tabular}
  \caption{Expected precisions of Higgs couplings, total Higgs width, and Higgs self-coupling for both baseline and luminosity upgrade (LumiUP) scenarios, at each running stage 250 GeV, 500 GeV, and 1 TeV, where the data at earlier stages is always combined to those at the current stage.}
\label{tab:HiggsCouplingsAll}
  \end{center}
\end{table}

\section{Summary and Acknowlegement}
The physics case at the ILC has a solid base complementary to the LHC, and provides a complete precision Higgs program without any model dependence, which is the key to reveal the secret of the EWSB and open the door to BSM physics. The capability of staged running is one great advantage and is definitely essential for accessing all the major Higgs couplings. Most importantly the ILC is technically ready to go! 

The authors would like to thank all the members of ILC physics subgroup and ILD optimization group for useful discussions, especially to the colleagues working on the Higgs analyses, T. Barklow, M. Berggren, C. Calancha, C. D$\ddot\mathrm{u}$rig, A. Ishikawa, S. Kawada, M. Kurata, A. Miyamoto, J. List, H. Ono, T. Price, T. Suehara, T. Tanabe, S. Watanuki, R. Yonamine. This work is supported in part by the Creative Scientic Research Grant No. 18GS0202 of the
Japan Society for Promotions of Science (JSPS), the JSPS Grant-in-Aid for Science Research No. 22244031, and the JSPS Specially Promoted Research No. 23000002.


\bibliographystyle{hunsrt}

\bibliography{ref}

\end{document}